%% file: template.tex
\documentclass[preprint, journal]{vgtc}       

\ifpdf
  \pdfoutput=1\relax                   
  \usepackage{graphicx}                
  \DeclareGraphicsExtensions{.pdf,.png,.jpg,.jpeg} 
\else
  \usepackage{graphicx}                
  \DeclareGraphicsExtensions{.eps}     
\fi%

\graphicspath{{figures/}{pictures/}{images/}{./}} 

\usepackage{microtype}                 
\PassOptionsToPackage{warn}{textcomp}  
\usepackage{textcomp}                  
\usepackage{mathptmx}                  
\usepackage{times}                     
\usepackage{cite}                      
\usepackage{tabu}                      
\usepackage{booktabs}                  

\usepackage{soul}

\usepackage{xcolor}
\usepackage{graphicx}
\usepackage{tabularx}
\usepackage{multirow}
\usepackage{fontawesome}
\usepackage{enumitem}
\usepackage{verbatim}

\newcommand{\emily}[1]{\textcolor{magenta}{#1}}
\newcommand{\ryan}[1]{\textcolor{blue}{#1}}
\newcommand{\arpit}[1]{\textcolor{violet}{#1}}
\newcommand{\ali}[1]{\textcolor{orange}{#1}}
\newcommand{\removed}[1]{\textcolor{red}{#1}}
\newcommand{\revised}[1]{\textcolor{black}{#1}}

\MakeRobust{\emily}
\MakeRobust{\ryan}
\MakeRobust{\arpit}
\MakeRobust{\ali}
\MakeRobust{\removed}

\newcommand{\system}{{\rmfamily \scshape vitaLITy}}
\newcommand{\attr}[1]{\textbf{#1}}

\renewcommand{\removed}[1]{\unskip}

\ieeedoi{10.1109/TVCG.2021.3114820}

\onlineid{0}

\vgtccategory{Research}
\vgtcpapertype{please specify}

\title{\system: Promoting Serendipitous Discovery of Academic Literature with Transformers \& Visual Analytics}

\author{Arpit Narechania, Alireza Karduni, Ryan Wesslen, and Emily Wall}
\authorfooter{
\item
 Arpit Narechania is with Georgia Tech. E-mail: arpitnarechania@gatech.edu.
\item
 Alireza Karduni and Ryan Wesslen are with UNC-Charlotte. E-mails: \{akarduni, rwesslen\}@uncc.edu.
\item
 Emily Wall is with Emory University. Email: emily.wall@emory.edu *work performed while at Northwestern University.
}

\shortauthortitle{Narechania \MakeLowercase{\textit{et al.}}: Word Embedding Approach for Identifying Relevant Vis Literature}

\abstract{There are a few prominent practices for conducting reviews of academic literature, including searching for specific keywords on Google Scholar or checking citations from some initial seed paper(s).  
These approaches serve a critical purpose for academic literature reviews, yet there remain challenges in identifying relevant literature when similar work may utilize different terminology (e.g., mixed-initiative visual analytics papers may not use the same terminology as papers on model-steering, yet the two topics are relevant to one another). 
In this paper, we introduce a system, \system, intended to complement existing practices. 
In particular, \system{} promotes serendipitous discovery of relevant literature using transformer language models, allowing users to find semantically similar papers in a word embedding space given (1) a list of input paper(s) or (2) a working abstract. 
\system{} visualizes this document-level embedding space in an interactive 2-D scatterplot using dimension reduction. 
\system{} also summarizes meta information about the document corpus or search query, including keywords and co-authors, and allows users to save and export papers for use in a literature review. 
We present qualitative findings from an evaluation of \system, suggesting it can be a promising complementary technique for conducting academic literature reviews.
Furthermore, we contribute data from 38 popular data visualization publication venues in \system, and we provide scrapers for the open-source community to continue to grow the list of supported venues.  
} 


\CCScatlist{ 
 \CCScat{K.6.1}{Management of Computing and Information Systems}%
{Project and People Management}{Life Cycle};
 \CCScat{K.7.m}{The Computing Profession}{Miscellaneous}{Ethics}
}

\teaser{
  \centering
  \includegraphics[width=\linewidth]{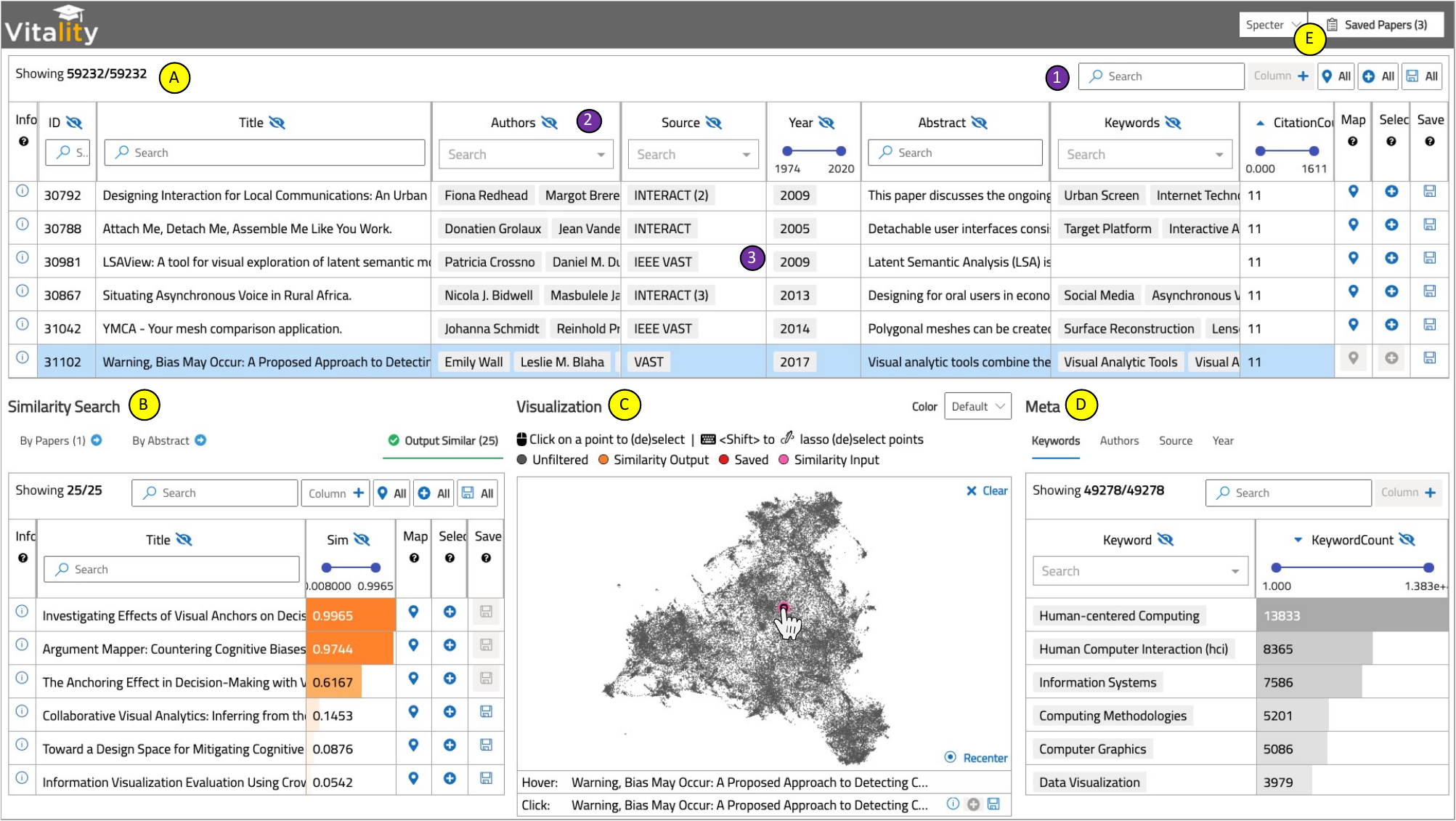}
  \caption{The \system{} User Interface. (A) \textbf{Paper Collection View} shows the entire corpus of publications, (B) \textbf{Similarity Search View} shows options to look-up publications that are similar to another list of publications or by a work-in-progress title and abstract, (C) \textbf{Visualization Canvas} shows an interactive 2-D UMAP projection of the embedding space of the entire paper collection, (D) \textbf{Meta View} shows summaries of certain attributes with respect to the Paper Collection View (A), (E) Opens a \textbf{Saved Papers View} from where the saved papers can be exported as JSON. Within the Paper Collection View (A), (1) shows an overview with global UI controls (e.g., filters), (2) shows attribute-level UI filters (range sliders, multiselect dropdowns), and (3) shows an interactive table of all publications.}
  \label{fig:teaser}
}

\vgtcinsertpkg

\begin{document}


\firstsection{Introduction}

\maketitle


Visualization research is inherently interdisciplinary, borne out of fields such as Computer Graphics and Human-Computer Interaction, with heavy influence from fields outside of computing such as Perceptual Psychology and Cognitive Science. 
Furthermore, visualization is applied to explore data and support data-driven decision making problems in domains ranging from enterprise analytics to medicine. 
As a result of the multi-faceted nature of the field, there may be parallel research efforts that can be difficult to become aware of, even with a comprehensive methodology for conducting literature reviews. 

One challenge of interdisciplinary research is when different fields use similar terminology to study different problems. 
For instance, \emph{transformer} in electronics refers to a device that transfers energy between circuits~\cite{kulkarni2017transformer}; while in computing, \emph{transformer} refers to a type of neural network based on attention mechanisms, commonly applied to unstructured text data~\cite{vaswani2017attention}.
As a result, keyword searches often yield irrelevant work. 
Further, sifting through all hits from a keyword search may still miss critical work. 
For instance, the recent wave of work on \emph{bias} in visualization (e.g.,~\cite{dimara2017attraction,dimara2019mitigating,WallBias,wall2018four,cho2017,valdez2018priming,wall2021lrg,narechania2021lumos}) seldom mentions \emph{uncertainty} (e.g.,~\cite{hullman2015hypothetical,hullman2019authors}).
Yet, as the seminal work on bias in Cognitive Science points out, bias emerges when people make decisions under uncertainty~\cite{Tversky1974}; hence, there is a critical need to examine uncertainty literature that may fundamentally address similar problems using different terminology.
As a result, conducting a simple keyword search for ``bias'' (i.e., matching tokens in a paper title or abstract) 
to identify relevant work may neglect pockets of influential research.
However, these challenges are not unique to data visualization research or even computing. 
They extend to virtually all interdisciplinary research. 

Current prevalent practices for conducting literature reviews tend to utilize two common search strategies: (1) keyword search and (2) examination of back-references from a snowballing set of seed papers, usually through searching Google Scholar or DBLP. 
These approaches can successfully identify a large number of relevant citations, but can suffer from at least two key limitations: thoroughness and efficiency. 
That is, they may fail to unearth related papers that use different terminology, and they require significant manual effort to gauge relevancy of potentially thousands of hits. 
In other words, a prominent challenge, then, in conducting literature reviews or surveys is to effectively identify research of significance to a given topic based on similarity of topics, irrespective of matching exact keywords.

To address these challenges, we introduce \system{}, an open-source visualization system designed to support a flexible exploration of research articles. 
Inspired by work on \emph{insight} in visualization (i.e., ``eureka'' or ``aha'' moments~\cite{chang2009defining}), we similarly aim to support \emph{serendipity} with \system, \revised{operationally intended to describe the goal that users may ``stumble upon'' relevant literature, when other search approaches might otherwise fail}. 
\revised{\system{} incorporates SPECTER \cite{cohan2020specter}, a state-of-the-art document-level contextual embedding model for scientific document recommendation.
Unlike many pre-trained language models that use a general corpus like Wikipedia or the Common Crawl \cite{mikolov2013efficient,pennington2014glove,devlin2018bert}, SPECTER was pre-trained on academic literature (sciBERT \cite{beltagy2019scibert}) and fine-tuned with citations which provides out-of-the-box state-of-the-art performance for academic literature recommendations and topic classification.}



In summary, this work presents the following contributions: 
\begin{enumerate}[nosep]
    \item results of a formative interview study in which visualization researchers identified key challenges in current literature review practices (Section~\ref{sec:formative}),
    \item a dataset of scraped metadata from \texttt{59,232} academic articles (\textbf{\url{https://figshare.com/articles/dataset/VitaLITy_A_Dataset_of_Academic_Articles/14329151}}~\cite{Narechania2021}, CC0 License), including paper titles, keywords, and abstracts from \texttt{38} popular venues for visualization research (Section~\ref{sec:data}),
    \item an open-source tool, \system{} (\textbf{\url{http://vitality-vis.github.io}}, MIT License), for supporting discovery of relevant articles while conducting literature reviews (Section~\ref{sec:system_overview}),
    \item usage scenarios describing potential workflows in which \system{} might be used in different ways to support serendipitous discovery of relevant academic literature 
    (Section~\ref{sec:case_study}), and 
    \item results of a summative evaluation of \system{} (Section~\ref{sec:evaluation}).
\end{enumerate}

\input{sections/related_work}
\input{sections/formative}
\input{sections/system}
\input{sections/bias}
\input{sections/evaluation}
\input{sections/discussion}
\input{sections/conclusion}


\bibliographystyle{abbrv-doi}

\bibliography{template}
\end{document}

%% file: sections/related_work.tex
\section{Related Work} 
\label{sec:related_work}

\subsection{Literature Review Methodologies}
Literature reviews and surveys are an essential part of scientific disciplines. 
They are broadly defined as systematic ways of collecting and synthesizing research on a specific topic \cite{baumeister1997writing,snyder2019literature}. 
There are a variety of different guidelines and methodologies, such as systematic reviews \cite{moher2009preferred}, narrative reviews \cite{baumeister1997writing}, and integrative reviews \cite{torraco2005writing}. 
These guidelines and methods mostly vary in how they organize, synthesize, and analyze a set of selected articles through a combination of quantitative and qualitative methods \cite{snyder2019literature}. 
These methodologies often include multiple stages, the first of which is related to identifying a strategy for searching and selecting a set of related literature. 
For example, Hannah Snyder states that \textit{``a search strategy for identifying relevant literature must be developed. 
This includes selecting search terms and appropriate databases and deciding on inclusion and exclusion criteria. 
Here, a number of important decisions must be made that are crucial and will eventually determine the quality and rigor of the review''}\cite{snyder2019literature}. 

Similarly, within the visualization community, defining search strategies and keywords are described as the primary step for conducting literature reviews \cite{mcnabb2019write}.
Many visualization survey papers include explicit excerpts about their selection criteria that describe keywords, databases, and the search process of each survey paper \cite{tong2018storytelling,fuchs2016systematic,roberts2018visualising}. 
For example, in their survey of glyph visualization techniques, Fusch et al. employ a ``snow ball'' sampling technique in which they start by searching the keyword ``glyph'' within various libraries, select all the findings, filter based on their exclusion criteria, and then look at the related work of the selected papers to find more papers \cite{fuchs2016systematic}. 

Although keyword search is the most prevalent method for searching literature, it comes with some limitations:
\begin{itemize}[nosep]
    \item Often it won't yield papers that do not include a specific keyword but might be very related to the topic at hand.
    \item Within different communities, different keywords are used to represent a common concept.
\end{itemize}

As a result, selecting sufficiently broad yet relevant keywords can be a challenge.
\system{} offers a visual system that complements traditional keyword search-based methods to enhance literature searches.
\system{} implements a state-of-the-art transformer-based document similarity search that can find semantically similar documents that may not always share the same set of keywords.




\subsection{Visualization of Academic Articles}
Visual analytics research has been effective in incorporating many machine learning and natural language processing models (e.g., topic modeling or word embeddings) into vis systems for exploratory analysis of large corpora of text documents \cite{Endert2012,liu2018bridging,dou2013hierarchicaltopics,el2017progressive}. A common task is identifying similar documents \cite{endert2017state}. Early visualization papers on document similarities used representations of a corpus' similarity matrix through dot plots \cite{church1993dotplot} or histograms \cite{freire2008visualizing}. More recent vis systems have considered more author assigned keyword-based approaches like constructive text similarity \cite{abdul2017constructive} and GlassViz \cite{benito2020glassviz}. 
Alternative approaches have considered word embeddings including for iterative lexicon construction \cite{park2017conceptvector} that provide related ability to query documents. 

One key application area for incorporating visual techniques to help users find similar and relevant documents is in searches for academic articles. Several prominent article databases have implemented such systems to find relevant articles. Text Analyzer by JSTOR extracts the most important topics and keywords from entered papers and recommends other relevant documents to users (\url{https://www.jstor.org/analyze/}). Pubmed uses a word-based technique to help users retrieve the most similar papers ( \url{https://pubmed.ncbi.nlm.nih.gov/help/#pubmedhelp.Computation_of_Weighted_Relev}). Open Knowledge graph uses similarity scores provided by Pubmed and develops a circle packing visualization to help users understand groups of related research relevant to their search terms (\url{https://openknowledgemaps.org/}). 

Within the visualization community, several works highlight the importance of understanding and visualizing academic literature. 
Felix et al., introduce a design space and highlight how different keyword summarization techniques might impact users' understanding of related literature \cite{felix2017taking}. 
Using the open source vis literature dataset (VisPubData), Isenberg et al. introduce KeyVis and analyze keywords utilized in the visualization community~\cite{isenberg2016visualization,Isenberg:2017:VMC}. 
Others introduce relevant systems for supporting dissemination of curated survey results~\cite{beck2015visual}, \revised{visualization of lead-lag analysis of text corpora~\cite{liu2014exploring}}, analysis of the contextual \emph{reasons} for citations~\cite{yoon2020conference}, and an emergent design space for considering visualizations of literature collections~\cite{hinrichs2015speculative}. 
In general, within visualization systems on academic literature we can observe three themes: (1) visualization systems that focus on citation networks (e.g.,~\cite{wilkins2015evolutionworks,heimerl2015citerivers,chou2011papervis,dattolo2018visualbib}), systems that focus on clustering or similarity (typically by matching keywords, e.g.,~\cite{wang2019vispubcompas}, or using topic modeling, e.g.,~\cite{alexander2014serendip,isenberg2016visualization}), and (3) systems that focus on both citation networks and similarity measures (e.g.,~\cite{nakazawa2018analytics,chen2006citespace}).
\revised{In the latter category, CiteSpace II introduces a technique to computationally define \emph{co-citation clusters}.}

Inspired by these works, our paper introduces (1) a more comprehensive public dataset of visualization literature, and (2) utilizes state-of-the-art document embedding techniques using transformers to enable serendipitous discovery of articles. 


%


\subsection{Word Embeddings and Transformers}
Document similarity is a classic problem in natural language processing and information retrieval \cite{Jurafsky2021}. 
Word embeddings provide an approach in which words (or documents) that have similar meanings have similar (vector) representations.
\revised{Recent advances in word embeddings have yielded significant improvements in standard similarity benchmarks like STS or SentEval \cite{arora2017simple,reimers-gurevych-2019-sentence,cohan2020specter}.} 
Beginning with word2vec \cite{mikolov2013efficient}, many extensions of learned dense representations of word vectors have followed including GloVe \cite{pennington2014glove}, fasttext \cite{bojanowski2017enriching}, skipthought \cite{kiros2015skip}, ELMo \cite{peters2018deep}, and BERT \cite{devlin2018bert}. \revised{More recently, specialized transformer models like SPECTER \cite{cohan2020specter} have been developed to specialize in domains like academic literature. SPECTER combines self-supervised pre-training on transformer architectures (e.g., BERT-like) on academic abstracts and is ``citation-informed'' to enhance performance for tasks like academic literature recommendation and topic classification.} 

\revised{SPECTER provides four advantages over past word embedding  approaches for \system{}'s task. First, it incorporates contextual embeddings (via BERT/transformer architecture) that enable different vector representations depending on the context (e.g., ``bias'' in different contexts). Second, the model was pre-trained on academic titles and abstracts (sciBERT \cite{beltagy2019scibert}). This enables the model to have transfer learning gains from pre-training with a BERT-like \cite{devlin2018bert} transformer architecture but with specialization for academic literature recommendation. Third, it incorporates a triplet-loss pre-training objective that enables it to use citations as an inter-document incidental supervision signal for fine-tuning. By incorporating both text pre-training with citation fine-tuning, the model achieved state-of-the-art performance for academic literature recommendations as well as six additional tasks like citation prediction, user activity (view or read), and topic classification. Tasks like citation prediction or user activity were out of scope of \system{}’s design due to data limitations, but future work could easily incorporate such tasks with additional citation or activity data. Fourth, the model is available out-of-the-box without fine-tuning as well as in model deployment through a publicly released API. This API enables fast and efficient real time scoring in \system{}.} 


%% file: sections/formative.tex
\section{Formative Study}
\label{sec:formative}

We conducted a formative study to better understand the needs of researchers as they perform literature reviews.  
Participants were 4 Computer Science PhD students (3 female, 1 male; avg. 2.75 yrs. into PhD program) who had prior experience conducting literature reviews in the field of visualization.
Sessions lasted approximately 45 minutes. 
Participation was voluntary with no compensation. 

We presented the first two participants with an initial version of the literature review tool. 
After incorporating feedback in the next iteration of the system, we worked with the next two participants using the updated system.
Finally, we incorporated feedback from all formative study participants in \system{}, presented in the next section.

\subsection{Current Workflow}
\label{sec:workflow}

After obtaining informed consent, we asked participants to describe their typical workflow for conducting literature reviews via a semi-structured interview.
Participants expressed some haphazard nature to the beginning of their processes, e.g., \textit{``someone tells [them] about a paper, and [they] look up the citations and branch out from there''} (P1) or \textit{``use a starting point from an advisor''} (P2).
From there, there are some commonalities in processes.

Participants all utilized keyword searches on Google Scholar (P1-4).
As a fairly comprehensive database, participants did not worry whether a venue or paper would be present, and they appreciated the ``cited by'' feature to identify more recent relevant papers.
However, participants also expressed that keyword searches on Google Scholar result in many irrelevant papers that require a lot of manual filtering for relevance.
For instance, P4 viewed Google Scholar as a last resort, expressing they really only use it \textit{``if [they] don't have a better starting point seed paper.''} 
Echoing some of the motivation for this work, P1 indicated, \textit{``if a keyword is used differently in different fields, [they] have to read a lot of abstracts to determine whether it's relevant or not.''}

While Google Scholar seems to be the default search tool, there are others that participants integrate in various parts of their workflow when conducting literature reviews. 
For instance, P4 indicated regular use of bibliography management tools like Mendeley and Zotero.
Among our relatively small sample in this formative study, participants did not mention some other elements in their workflow that we anticipated, e.g., DBLP, manual scripting / web scraping, etc.

\subsection{Preliminary Feedback}
\label{sec:prelim_feedback}

Next, participants used a preliminary version of our literature review tool. 
The preliminary tool included \texttt{17,926} papers from the following venues over the past \texttt{39} years (1982-2020): \{\emph{CGA, CGF, EuroVis, Graphics Interface, Information Visualization, Interact, Journal of Visualization, PacificVis, SciVis, TVCG, VAST, VIS}\} and supported two main mechanisms for searching the corpus: keyword search and similarity search (described in greater detail in the next section).
After using the tool, we asked participants for additional feedback about the current implementation, possible improvements, and any new capabilities that they could envision to better support their literature review process. 
 
Participants appreciated the ability to start their search with a seed paper or papers (P1 said they got \textit{``pages and pages of results which is what [they] would get on Google Scholar, but these are actually more relevant''}). 
P2 searched based on the seminal paper on hypothetical outcome plots (HOPs)~\cite{hullman2015hypothetical} and observed \textit{``it pulled up lots of uncertainty vis papers, which were not in the title -- cool!''}
but expressed that there was still a lot of noise when searching by keywords.

Participants \revised{suggested several} new features: being able to visualize connections between papers (e.g., by citations, co-authors, etc. - P1), adding critical information on citation count as a mechanism for determining importance of a paper (P1), making the overview interactive (with brushing and linking, summarizing dynamic regions, etc. - P2), and being able to type in a custom abstract or paper idea as the basis of the similarity search (e.g., to identify relevant literature for a paper idea that hasn't been fully fleshed out yet - P2). 
Participants also steered away from one of the features in the tool: the word cloud. 
P4 indicated \textit{``it wasn't clear how it was related to what [they] had selected.''}

Overall, participants indicated that a tool like this in their workflow could supplement tools like Google Scholar for serendipitous exploration. 
P4 suggested it would be beneficial in the early ``discovery'' phases of literature review, with the caveat that the data on included venues needed to be sufficiently comprehensive. 
As a result of this feedback, we updated the system to address these ideas, including scraping data from additional venues, adding citation counts, adding brushing and linking between views, and searching by a custom abstract. 
We did not add features based on citation networks in our system\revised{; instead, we focused on leveraging transformer models to serve as a complementary literature search technique to existing tools that address these needs}. \removed{due to space and complexity constraints in the tool.}

\subsection{Design Goals}
\label{sec:design_goals}

Collectively, these interviews led us to the following set of four design goals for our literature review system.

\smallskip
\noindent\textbf{DG 1. Serendipity:} Enable serendipitous identification of semantically related articles that do not necessarily have shared keywords through visual exploration.

\smallskip
\noindent\textbf{DG 2. Familiarity:} Facilitate a familiar search functionality to what users are currently accustomed to, such as keyword and author search. 

\smallskip
\noindent\textbf{DG 3. Novelty:} Afford users to find semantically related articles by searching based on the author's own ideas in the form of unpublished sentences / abstract.

\smallskip
\noindent\textbf{DG 4. Overview:} Enable users to interact with a visual overview of a group of papers.


%% file: sections/system.tex
\section{\system}
\label{sec:system}

\begin{figure*}[!t]
  \centering
  \setlength{\belowcaptionskip}{-10pt}
  \includegraphics[width=\linewidth]{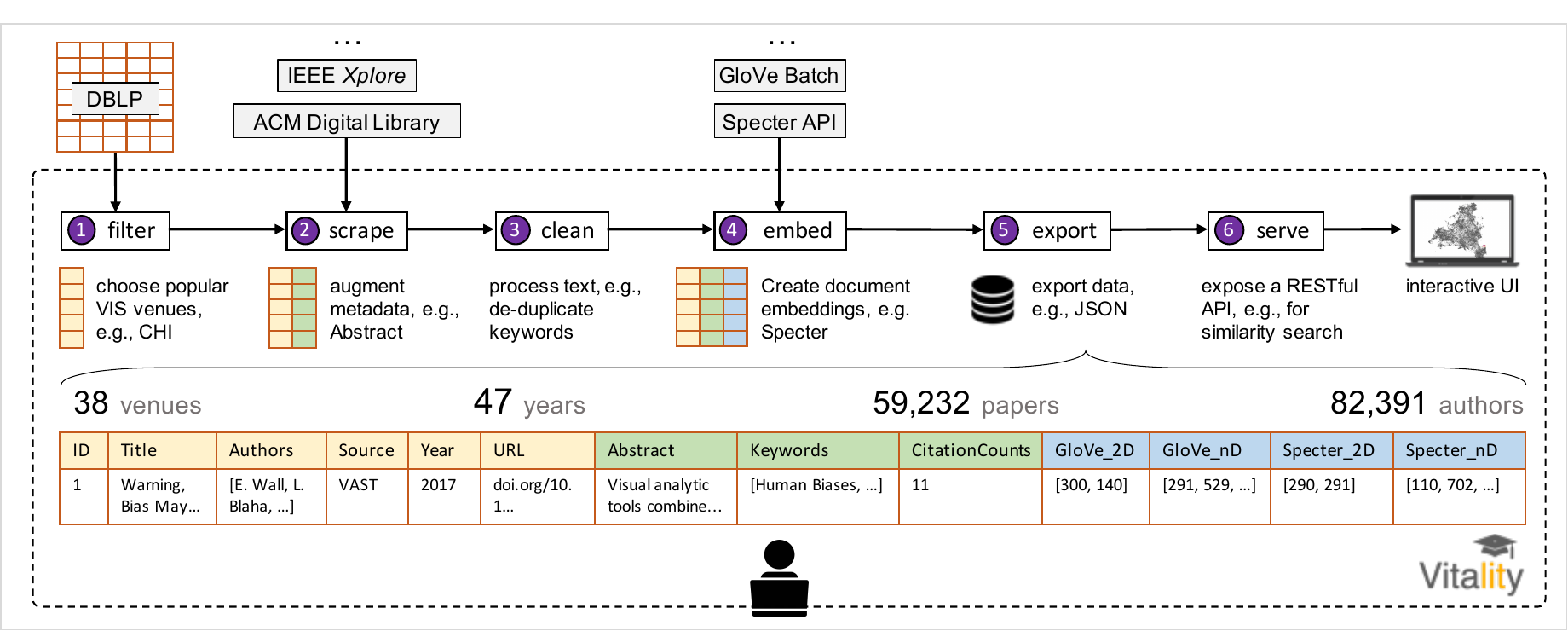}
  \caption{The \system{} architecture. (1) DBLP data is filtered by relevant venues. (2) Author and title metadata from DBLP is augmented with abstracts, keywords, and citations from custom scrapers. (3) Data is cleaned (e.g., to resolve duplicate keywords, etc). (4) GloVe and Specter document embeddings are created. (5) Data is exported to a variety of formats for subsequent open-source use. (6) The server exposes a RESTful API that can ultimately be called upon in rendering the interactive system.}
  \label{fig:architecture}
\end{figure*}

We present \system{}, a system designed to complement existing tooling for conducting academic literature reviews by supporting serendipitous discovery of relevant literature. 

\subsection{Data} 
\label{sec:data}
Figure~\ref{fig:architecture} outlines the pipeline for curating the paper corpus.

\begin{figure}[t]
    \centering
    \includegraphics[width=\linewidth]{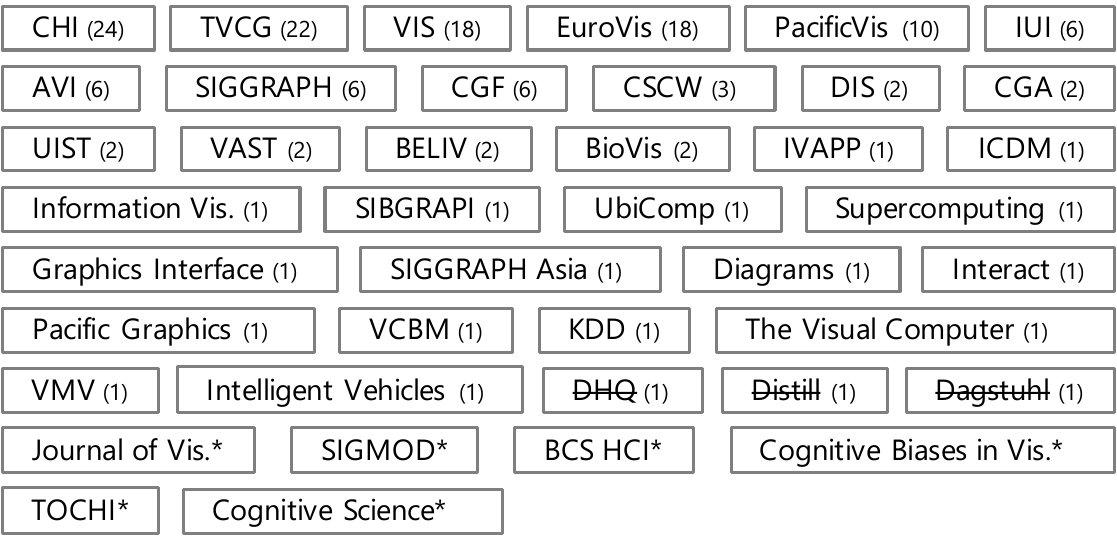}
    \caption{Results from a Twitter survey with \texttt{24} users on venues where VIS researchers publish; (numbers in parentheses) are aggregated counts of the \texttt{24} responses; \st{struckthrough venues} were not in DBLP and hence currently not available in \system{}; * venues were added as also-relevant venues by the authors after the survey; ``Vis.'' is short for Visualization.}
    \label{fig:venues}
\end{figure}

\noindent\textbf{1. Filter:} We conducted an open-ended crowd-sourced survey on Twitter asking visualization researchers about venues (e.g., journals, conferences, workshops) where they publish. We received responses from \texttt{24} users (current roles: 17 Ph.D. students, 4 Faculty, 2 Industry researchers, and 1 Postdoctoral scholar; self-reported visualization literacy out of 5: $\mu$=4, $\sigma$=1.142, median=4). We supplemented the list with \texttt{six} additional venues based on our own knowledge that were not captured in the survey. Figure~\ref{fig:venues} outlines the final list of \texttt{38} venues within our corpus. Next, we downloaded the November 2020 release~\footnote{\url{https://dblp.org/xml/release/dblp-2020-11-01.xml.gz}} of DBLP~\cite{dblp} and filtered it for the aforementioned \texttt{38} venues. From the resultant subset, we chose the \attr{Title}, \attr{Authors}, \attr{Source} (venue), \attr{Year} (published), and \attr{URL} attributes and added a unique \attr{ID} for tracking. Note that the \system{} dataset and hence the UI show more than 38 venues because DBLP (a) utilizes multiple descriptors to represent different tracks at the same venue (e.g., Eurographics (Area Papers), Eurographics (State of the Art Papers), Eurographics (Short Papers), etc.) and (b) splits some venues across different versions (e.g., Interact, Interact (1), Interact (2), etc.).

\noindent\textbf{2. Scrape:} The DBLP dataset does not include abstracts, keywords, and number of citations for papers. Thus, we developed a \emph{scraper} module that, given a list of publication URLs, scrapes the corresponding publisher's webpage (e.g., IEEE Xplore, ACM Digital Library) and extracts the \attr{Abstract}, \attr{Keywords}, and \attr{CitationCounts} from it. 

\noindent\textbf{3. Clean:} We performed data cleaning and transformation operations. To aid search, we encoded all text attributes to ASCII and converted \attr{Authors} and \attr{Keywords} into a JSON array from a comma separated list. We de-duplicated \attr{Keywords} by matching their lowercase forms; we combined similar keywords (e.g., HCI \& Human-Computer Interaction, Visuali\textbf{z}ation \& Visuali\textbf{s}ation) through manual inspection. We dropped \texttt{1497} papers with \emph{null} \{\attr{Title}, \attr{Authors}, \attr{Abstract}\} values, and very short or very long \attr{Title} ($<$5, $>$250 characters) and \attr{Abstract} ($<$50, $>$2500 characters) to create effective word embeddings. We retained DBLP's strategy in disambiguating author names (e.g., \emph{J. Thompson} and \emph{J. Thompson 001}). At the end of this step, the dataset has \texttt{8} attributes (columns) and \texttt{59,232} papers (rows).

\noindent\textbf{4. Embed:} We next curated a dataframe of \attr{Title}, \attr{Abstract}, \attr{Authors}, \attr{Source}, \attr{Year}, and \attr{Keywords} and created the GloVe \cite{pennington2014glove} and Specter\cite{cohan2020specter} document embeddings. To create the document embeddings for GloVe, we used TF-IDF weightings (instead of mean vectors) and SIF weightings that have been shown to remove noise through PCA reduction \cite{arora2017simple}. We used the public API to create the Specter embeddings~\cite{cohan2020specter}. With these document embeddings, we used UMAP to construct 2-D document representations used in the Visualization Canvas (see Figure \ref{fig:umap}).

\noindent\textbf{5. Export:} We export the consolidated dataset as JSON and a MongoDB dump for different open-source use.

\noindent\textbf{6. Serve:} We also developed a server that exposes a RESTful API to (a) load the \system{} document corpus, (b) perform similarity search by a list of seed papers as input, (c) perform similarity search by a working title and abstract as input, and (d) download metadata of (saved) papers as a JSON array. The similarity search by seed papers (b) supports querying by 2-D UMAP as well as n-D document embeddings for both GloVe and Specter. We used MongoDB to maintain the 2-D indexes and Facebook Research's faiss library~\cite{faiss} to maintain the n-D indexes. \revised{For one seed paper as input, we utilize existing APIs to compute the Euclidean (2-D; MongoDB) and L2 (n-D; Faiss) distances between the input paper and other papers, compute their reciprocals, and normalize them between 0-1 for use as the similarity scores (1 = most similar). For more than one seed paper as input, we first compute the average vector from all input papers and then follow the same procedure as above to compute the similarity scores. 
} The \system{} UI interfaces with this server, described next.

\subsection{System Overview}
\label{sec:system_overview}

The system, shown in Figure~\ref{fig:teaser}, is comprised of a \emph{Paper Collection View} (A), \emph{Similarity Search View} (B), \emph{Visualization Canvas} (C), \emph{Meta View} (D), and \emph{Saved Papers View} (E), described in turn below.

\smallskip
\noindent\textbf{Paper Collection View} shows the entire corpus of papers in an interactive tabular layout. (1) shows an overview (number of visible papers) and UI controls to perform a global search (\faSearch), show hidden columns ([Column~\faPlus]), add all papers to the input list of papers in the \emph{Similarity Search} table ([\faPlusCircle~All]), and save all papers to the ``cart'' in the \emph{Saved Papers View} ([\faSave~All]). (2) shows the attributes along with UI controls to filter (range sliders for Quantitative attributes, multiselect dropdowns for Nominal attributes), hide a column (\faEyeSlash), and define a column on hover (\faQuestionCircle). (3) shows an interactive table of all papers with options to see detail 
(\faInfoCircle), 
locate in the UMAP (\faMapMarker), add to the input list of papers for similarity search (\faPlusCircle), and save to the ``cart'' (\faSave). Search and filter capabilities are designed to be an intuitive entry-point into the dataset of academic articles (\textbf{DG 2}).

\smallskip
\noindent\textbf{Similarity Search View} shows options to find papers similar to (a) one or more input papers (Figure~\ref{fig:search-by-papers}, \textbf{DG 1}) or (b) a work-in-progress title and abstract (Figure~\ref{fig:search-by-abstract}, \textbf{DG 3}). \system{} supports setting the dimensions (2-Dimensional, n-Dimensional), number of similar papers to return, and the word embedding approach (e.g., Specter) to compute similarity.

\begin{figure}[!t]
    \centering
    \setlength{\belowcaptionskip}{-10pt}
    \includegraphics[width=\columnwidth]{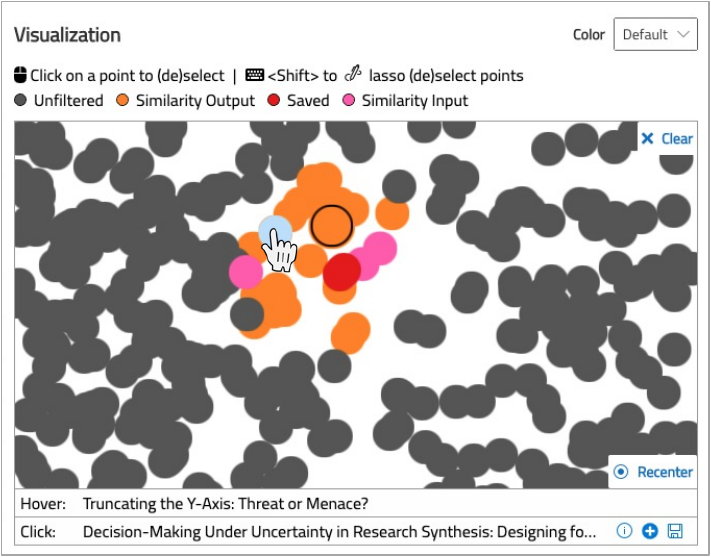}
    \caption{Interactive 2-D scatterplot of the UMAP projection.}
    \label{fig:umap}
  \end{figure}

\smallskip
\noindent\textbf{Visualization Canvas} shows a 2-D UMAP projection of the embedding space of the entire paper collection (Figure~\ref{fig:umap}, \textbf{DG 4}): hovering on a point highlights it, shows the corresponding title in a fixed tooltip below, and automatically scrolls the collection (table) to bring the corresponding paper (row) into the viewport; clicking on a point (de)selects it and shows it in the tooltip below with additional options to \faInfoCircle, \faPlusCircle, \faSave; clicking on \faTimes~deselects all selected points; pressing Shift enables lasso-mode to select multiple points using a free-form lasso operation; zooming and panning support helps navigate the UMAP to specific regions; clicking on \faDotCircleO~re-centers and fits the UMAP in the viewport. By default, each point in the UMAP is colored based on the state of the corresponding paper (``Default''): Unfiltered (unfiltered and visible in the main paper collection table; dark-grey), Filtered (filtered out and not visible in the paper collection table; light-grey), Similarity Input (added to the \emph{By Papers} section in the Similarity Search View; pink), Similarity Output (in the \emph{Output Similar} table; orange), and Saved (added to the Saved Papers table; red). Other options to color include \attr{Source}, \attr{Year}, \attr{CitationCounts}, and \attr{Similarity Score}.

\begin{figure*}[!t]
    \centering
    \setlength{\belowcaptionskip}{-10pt}
    \includegraphics[width=\linewidth]{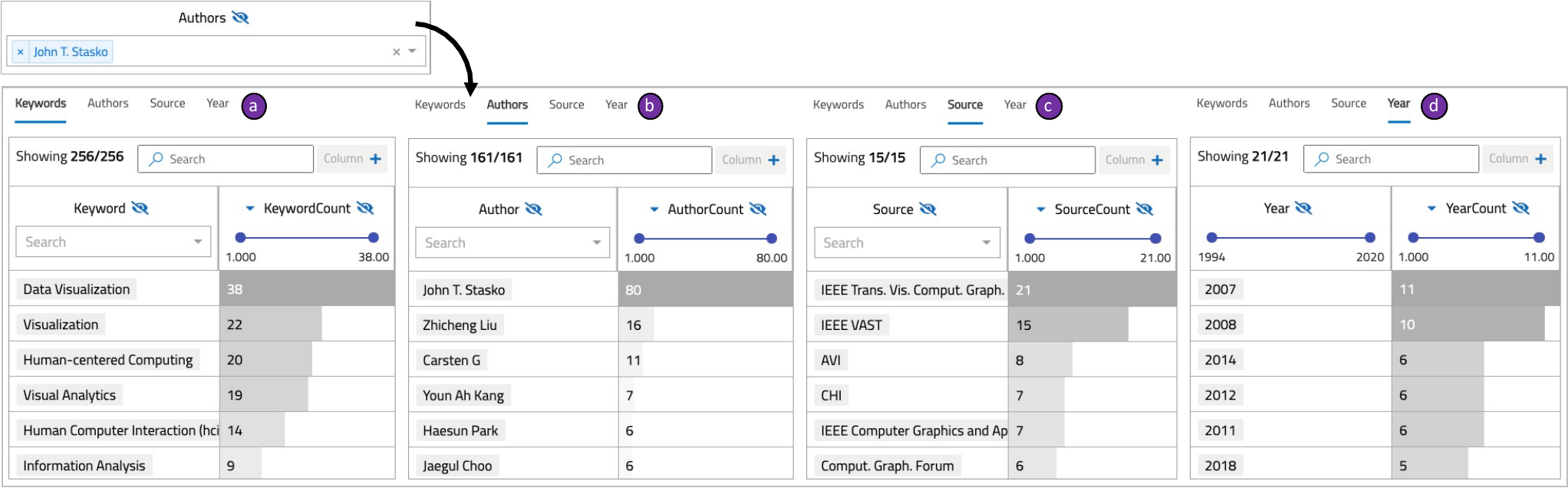}
    \caption{The Meta View showing aggregated summaries of (a) \attr{Keywords}, (b) \attr{Co-authors}, (c) \attr{Source}, and (d) \attr{Year} associated with \emph{John T. Stasko}.}
    \label{fig:meta}
  \end{figure*}

\smallskip
\noindent\textbf{Meta View} shows aggregated summaries of \attr{Keywords}, \attr{Authors}, \attr{Source}, \attr{Year} with respect to the \emph{Paper Collection View} (A). Figure~\ref{fig:meta} shows how a filter in the main table (\attr{Authors}=\emph{John T. Stasko}) updates the Meta Views with the distribution of keywords (a) associated with their research, their co-authors (b), venues where they have published (c), and in which years (d).

\smallskip
\noindent\textbf{Saved Papers View} shows a table with the papers added to the ``cart'' with an additional option to export them as a JSON (Figure~\ref{fig:search-by-papers}d).

\subsection{Implementation}
The \emph{filter}, \emph{scrape}, \emph{clean}, \emph{embed}, \emph{export}, and \emph{serve} modules are all implemented in Python. The \emph{UI} is developed in React and uses the regl-based WebGL library\footnote{https://github.com/flekschas/regl-scatterplot} to render the UMAP. MongoDB provides the document corpus to the UI and maintains the 2-D indexes while faiss~\cite{faiss} maintains the n-D indexes for efficient similarity search.

%% file: sections/bias.tex
\section{Usage Scenarios}
\label{sec:case_study}

A common thread among the authors' prior research deals with \textbf{human bias in data visualization}, and in particular, the authors have focused on defining~\cite{wall2018four}, detecting~\cite{cho2017,WallBias,WallFormative,WallMarkov}, and mitigating~\cite{WallDesign,narechania2021lumos,wall2021lrg} cognitive biases.
Thus, we find it fitting to demonstrate the usage of \system{} through a series of usage scenarios in the context of a literature review on bias in visualization.

\subsection{Usage Scenario 1: Identifying Missing Papers}
Maya is a data visualization PhD student working on their dissertation on the topic of ``Mitigating Bias in Data Visualization.''  
They are wrapping up the related work and preparing to submit their thesis. 
Before submitting, Maya wants to check for potential gaps in the literature review and ensure there is no critical missing work. 
Maya decides to use \system{} to explore the visualization literature.

\begin{figure*}[!t]
    \centering
    \setlength{\belowcaptionskip}{-10pt}
    \includegraphics[width=\linewidth]{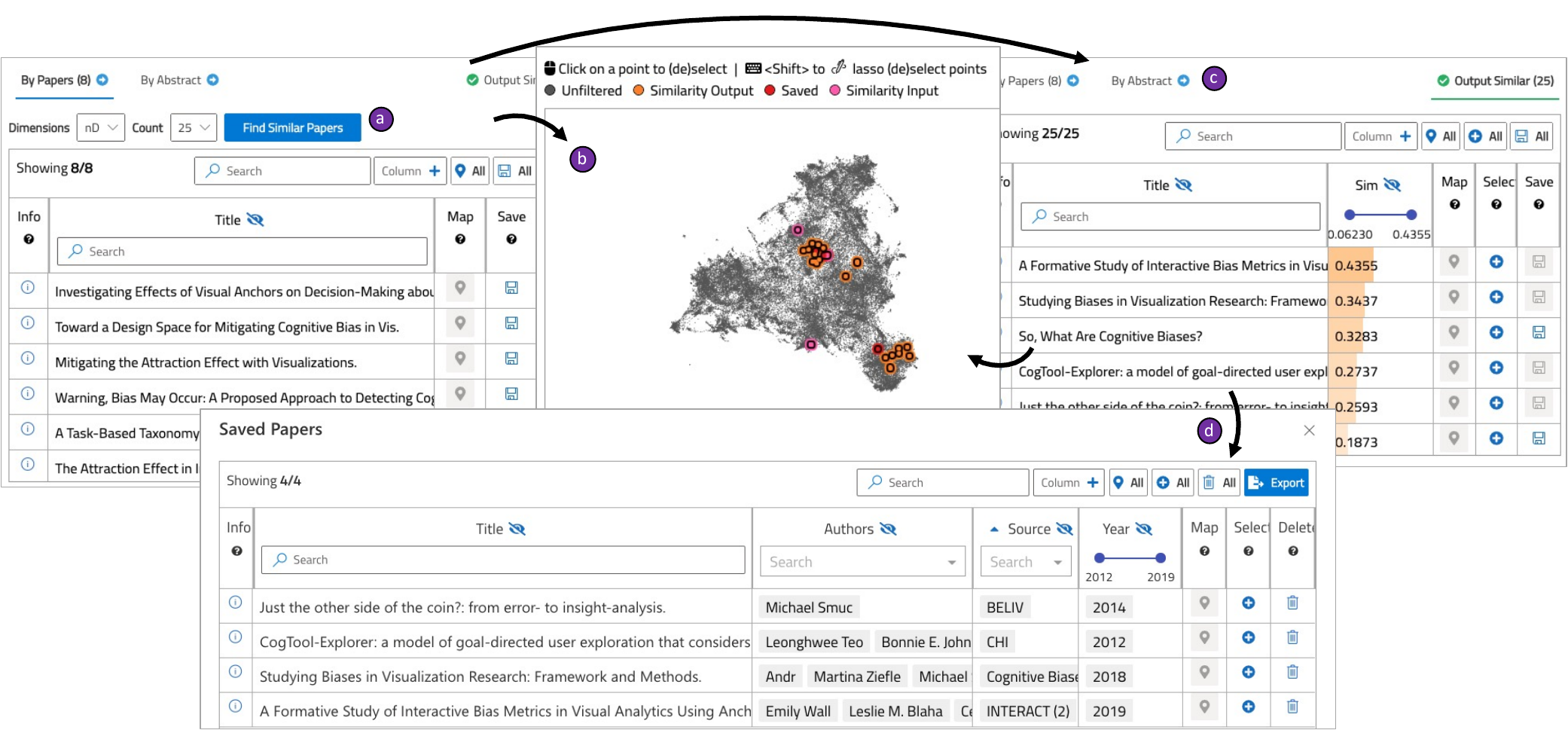}
    \caption{\textbf{Search by a list of seed papers: Scenario 1}. Based on a list of known relevant bias papers (a), Maya observes the clustering of similar papers in the Visualization (b). She examines the similar papers more closely to gauge their relevance (c) and exports relevant saved papers (d).}
    \label{fig:search-by-papers}
  \end{figure*}

Maya wants to be systematic about their search. 
They begin by taking some of the key papers related to bias in visualization, including the following~\cite{dimara2017attraction,dimara2018task,dimara2019mitigating,WallBias,WallDesign,cho2017,wesslen2019investigating,gotz2016adaptive}. 
Maya has already examined the papers cited from these works and written about the relevant ones in their dissertation. They locate these key papers in \system{} and ``select'' them [add them as input to Similarity Search] (Figure~\ref{fig:search-by-papers}a), then map them in the Visualization (Figure~\ref{fig:search-by-papers}b).

Starting with N-Dimensional Specter embedding, Maya searches for similar papers (Figure~\ref{fig:search-by-papers}c). 
The first result, ``A Formative Study of Interactive Bias Metrics in Visual Analytics Using Anchoring Bias''~\cite{WallFormative} (similarity score \texttt{0.4355}), is cited in one of the papers~\cite{WallDesign} so Maya was already aware. 
Scanning down the list, the fourth result is ``CogTool-Explorer: A Model of Goal-Directed User Exploration That Considers Information Layout''~\cite{teo2012cogtool}, a paper Maya is not aware of. 
Published at CHI in 2012, this paper describes a method for modeling and predicting user interactive behavior. 
Intrigued by the relevance of precursory work in HCI to predict interactive behavior~\cite{teo2012cogtool} to work on modeling user bias~\cite{WallBias}, Maya saves this paper to the ``cart''. 

Continuing to examine the list of output papers, the next result also proves relevant with a similarity score of \texttt{0.2593}: a BELIV paper titled ``Just the Other Side of the Coin? From Error to Insight Analysis''~\cite{smuc2016just} which models errors and insights in cognitive processing. 
Several others also catch Maya's eye relevant to the design of bias mitigation strategies, including research about introducing visualization ``difficulties'' in design to aid comprehension and recall~\cite{hullman2011benefitting} and even use of so-called ``transparent deception'' in visualization if and when it is aligned with certain user goals~\cite{ritchie2019lie}.
Maya saves these papers and exports them for further review (Figure~\ref{fig:search-by-papers}d). 

Furthermore, Maya notices a particularly relevant paper, ``Priming and Anchoring Effects in Visualization''~\cite{valdez2018priming}, which they forgot about, so adds it to the input similarity search and re-computes the output. 
They find ``Pushing the (Visual) Narrative: the Effects of Prior Knowledge Elicitation in Provocative Topics''~\cite{heyer2020pushing}, discussing persuasive visualization designs, which again Maya finds relevant for designing bias mitigation interventions. 
Maya continues iterating on their exploration of the literature, augmenting their dissertation related work section and filling in gaps, especially from the CHI community.



\subsection{Usage Scenario 2: Analysis of Keyword Quality}
Katherine is a visualization researcher who focuses on topics related to bias and decision making. 
She has primarily relied on keyword searches supported by IEEE Xplore, ACM Digital Library, etc. for identifying relevant literature in the past. 
Beginning with a set of known papers about bias in visualization (i.e., the same set from the previous scenario~\cite{gotz2016adaptive,WallBias,WallDesign,cho2017,wesslen2019investigating,dimara2017attraction,dimara2018task,dimara2019mitigating}), she identifies several relevant keywords, including
\emph{human biases}, \emph{bias mitigation}, \emph{bias mitigation strategies}, \emph{bias alleviation}, \emph{debiasing}, \emph{cognition}, \emph{cognitive bias}, \emph{cognitive biases}, \emph{cognitive heuristics}, \emph{heuristics},  \emph{decision making}, \emph{decision-making}, \emph{human decision-making}, \emph{sensemaking capabilities}, \emph{uncertainty}, \emph{anchoring bias}, and \emph{attraction effect}.
She disregards several others that she believes are too broad, e.g., \emph{visualization}, \emph{information visualization}, \emph{data visualization}, \emph{human-centered computing}, \emph{visual analytics}, etc. 
She notes the multiplicity of some keywords defined by authors. 

Katherine conducts a similarity search using \system{} (yielding the same output as the previous scenario for Maya's literature review). 
She notes a number of papers that she would have been unable to identify given only these keyword searches. 
For instance, ``Designing Information for Remediating Cognitive Biases in Decision-Making''~\cite{zhang2015designing} contains keywords \emph{Human Computer Interaction (hci)} and \emph{Human-centered Computing} and would have been missed by targeted bias-related keywords and likely lost among a sea of other papers by searching for more generic HCI keywords.
Similarly, ``A Lie Reveals the Truth: Quasimodes for Task-Aligned Data Presentation''~\cite{ritchie2019lie} contains very broad keywords, including \emph{Visualization}, \emph{Empirical Studies In Visualization}, and \emph{Human-centered Computing}.
Other papers not directly related to bias, but still relevant, are even less likely to contain keyword matches. 
For instance, ``Observation-Level Interaction with Statistical Models for Visual Analytics''~\cite{Endert2011} describes data- or ``observation''-level interactions users perform with data based on perceived relationships and interests in the data, a topic of precursory relevance to bias research in data visualization. 
However, it contains keywords with no overlap to the bias-related search terms: \emph{Principal Component Analysis}, \emph{Data Models}, \emph{Data Visualization}, \emph{Visual Analytics}, \emph{Analytical Models}, and \emph{Layout}.

Notably, Katherine observes that some venues expose only index terms from e.g., IEEE or ACM, while others also expose author-defined keywords. 
This provides different levels of granularity in the ability to search for literature by keyword. 
Hence, Katherine finds that alternative approaches based on document-level embeddings can be a fruitful way to identify literature when keyword searches prove insufficient or inconsistent across venues.

\subsection{Usage Scenario 3: Beginning a New Project} 

\begin{figure}[!t]
    \centering
    \includegraphics[width=\linewidth]{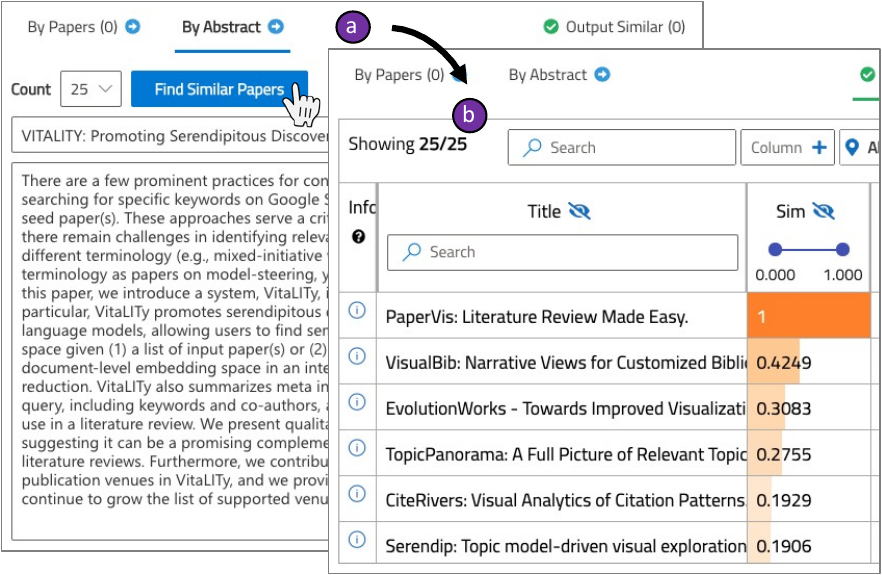}
    \caption{\textbf{Search by Abstract: \system{}'s own Literature Review.} The authors using \system{}'s working title and abstract (a) to find similar papers (b) to assist in its own literature review.}
    \label{fig:search-by-abstract}
  \end{figure}

In this scenario, we showcase how \system{} facilitated our own literature review for the present work. 
After using traditional approaches based on keyword searches or citations from known papers, we found \system{} helped us identify a plethora of additional literature we were previously unaware of.
We used the Similarity Search by Abstract feature of \system{} with our paper title and abstract (Figure~\ref{fig:search-by-abstract}a). 

The first returned result is a 2011 Computer Graphics Forum paper titled ``PaperVis: Literature Review Made Easy''~\cite{chou2011papervis} that utilizes a node-link visualization approach to support literature review and creates a topic hierarchy based on semantically meaningful topics (Figure~\ref{fig:search-by-abstract}b). 
The next paper similarly focuses on creating iterative citation networks to facilitate creation and sharing of bibliographies~\cite{dattolo2018visualbib}.
In general, after searching the output, a few themes emerge: (1) visualization systems that focus on citation networks (e.g.,~\cite{wilkins2015evolutionworks,heimerl2015citerivers}), systems that focus on clustering or similarity (typically by matching keywords, e.g.,~\cite{wang2019vispubcompas}, or using topic modeling, e.g.,~\cite{alexander2014serendip}), and (3) systems that focus on both citation networks and similarity measures (e.g.,~\cite{nakazawa2018analytics}).
Other notable topics also surfaced, including a design space~\cite{felix2017taking} and analyses of keywords utilized in the visualization community~\cite{isenberg2016visualization}, a system for supporting dissemination of curated survey results~\cite{beck2015visual}, analysis of the contextual \emph{reasons} for citations~\cite{yoon2020conference}, and an emergent design space for considering visualizations of literature collections~\cite{hinrichs2015speculative}.

Reflecting on these findings, we believe traditional methods for searching literature left many gaps in our literature review for two primary reasons: (1) many of these works are distributed across several publication venues (e.g., IV, PacificVis, Interact, VAST, TVCG), and (2) many of these papers received relatively little traction since their original publication 5-10 years ago.


\subsection{Usage Scenario 4: Getting to Know VIS}
Rosa is a new PhD student joining a lab that conducts research in data visualization. 
To become acquainted with the field, her advisor suggests that Rosa browse through some of the prominent literature in \system.
Upon loading the system, Rosa observes that it contains \texttt{59,232} papers in the Paper Collection View. 
Inspecting the Meta View, she observes those papers are described by \texttt{49,278} keywords, written by \texttt{82,391} authors from \texttt{55} different venues, across \texttt{47} years.
Among the top keywords are \emph{Human-centered Computing} and \emph{Human Computer Interaction (hci)}, describing \texttt{13,833} and \texttt{8,365} papers respectively. 
The lineage of data visualization becomes apparent to Rosa when she notices that the fifth most common keyword is \emph{Computer Graphics}, followed by \emph{Data Visualization}. 
Other common keywords that catch Rosa's eye describe topics such as \emph{Machine Learning}, \emph{Information Retrieval}, \emph{Artificial Intelligence}, \emph{Interaction Design}, and \emph{Animation}, among others. 

Rosa enters \emph{Data Visualization} as a filter in the Keywords column of the Paper Collection View, then filters to show only papers in the past 10 years to focus on the \texttt{2,032} most relevant recent works in the field. 
Interestingly, these papers appear in a fairly dense area near the center of the Visualization. 
In the Meta View, she notes a few authors whose names she recognizes, including Kwan-liu Ma who authored \texttt{58} of the papers with the keyword \emph{Data Visualization} since 2010. 
She also notices Daniel Keim, John T. Stasko, Niklas Elmqvist, and Hanspeter Pfister, among others. 
She next filters the Paper Collection View to see only John T. Stasko's papers (\texttt{80}) and removes the other filters (Figure~\ref{fig:meta}a-d). 
The Meta View reveals that his work is associated with the following keywords: \emph{data visualization, visualization, human-center computing, visual analytics, human computer interaction (hci)} (a). Some of his common co-authors include Zhicheng Liu, Carsten Gorg, and Youn Ah Kang (b).
He publishes primarily at TVCG (\texttt{21}) and VAST (\texttt{15}) (c), with 2007 his most productive year (\texttt{11} publications) followed by 2008 (\texttt{10} publications) then 2011, 2012, and 2014 each with \texttt{6} publications (d). 



%% file: sections/evaluation.tex
\section{Evaluation}
\label{sec:evaluation}

Based on the final form of \system{}, developed from formative feedback with visualization researchers, we next describe the summative evaluation of \system{} in a qualitative study.
We recruited 6 Computer Science PhD students (1 female, 5 male; avg. 3.3 yrs. into PhD program) whose primary research area is within the field of visualization.
None of the participants were involved in the formative study. 
Sessions lasted approximately 45 minutes.
Participation was voluntary with no compensation.

\subsection{Task \& Procedures}
\label{sec:task}
After obtaining informed consent,
participants were asked to reflect on a topic for which they had recently or were currently conducting a literature review. 
\system{} was loaded with all \texttt{59,232} papers, described in Section~\ref{sec:data}, running locally on the study investigator's machine. 
Participants connected to the study virtually via Microsoft Teams. 
They were asked to \textbf{recreate or continue their literature review using \system}, which they interacted with by using Microsoft Teams's ``Request Control'' feature on the study investigator's machine.
We utilized a think aloud protocol to capture users' impressions and qualitative feedback on the system. 
Sessions were screen-recorded for subsequent analysis.

Participants chose the following topics for their literature reviews: \emph{multiple comparisons problem}, \emph{interpretable machine learning}, \emph{misinformation}, \emph{network visualization}, \emph{scrollytelling visualization}, and \emph{transformation / similarity between two visualizations}.

\subsection{Findings}
\label{sec:findings}
\begin{figure}[!t]
    \centering
    \setlength{\abovecaptionskip}{4pt}
    \includegraphics[width=\columnwidth]{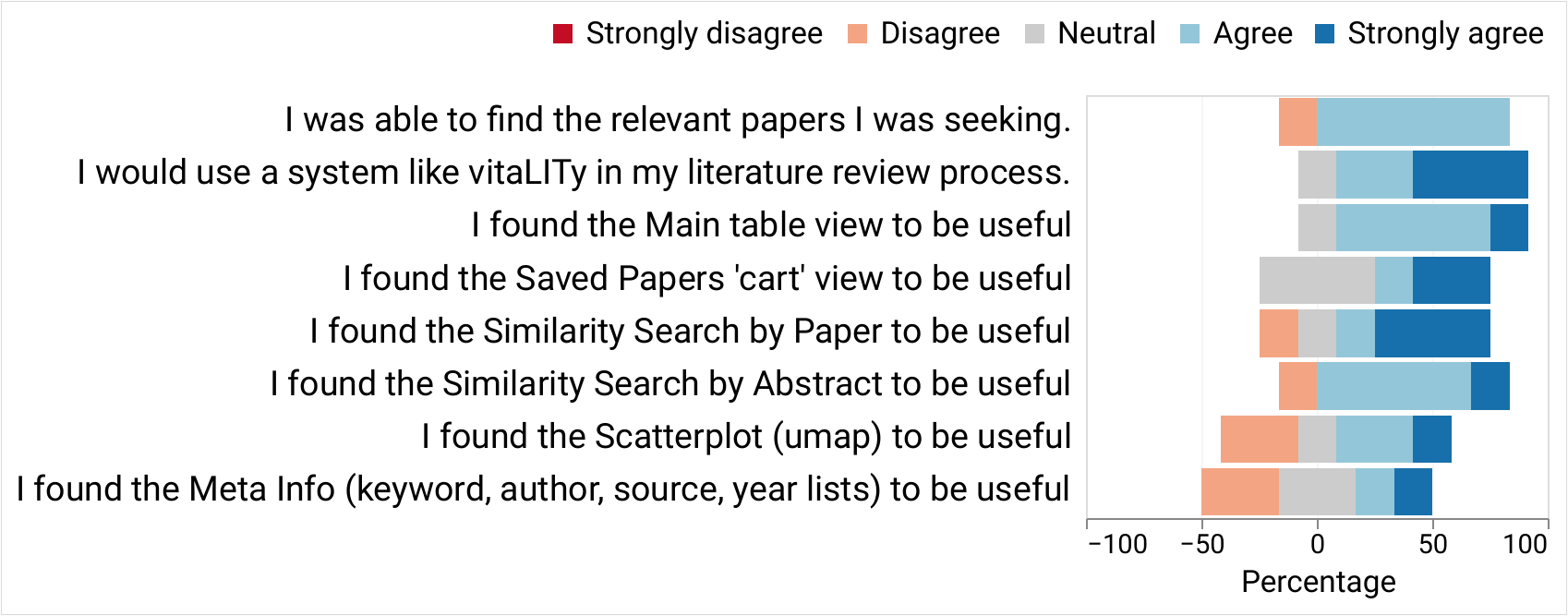}
    \caption{Usability scores of \system{} features.}
    \label{fig:sus-scores-specific}
  \end{figure}

In this section, we discuss qualitative findings from our evaluation of \system{} for each of its primary features (Figure~\ref{fig:sus-scores-specific}), and lastly summarize participants' general impressions of the system (Figure~\ref{fig:sus-scores-generic}).

\subsubsection{Paper Collection View}
Participants felt that the Paper Collection View was a good ``entry point'' into the paper corpus in \system, containing familiar data that users expected to see, e.g., authors, abstracts, etc (S02). 
Searching by keyword was familiar and produced expected outcomes.
For instance, S03 identified some papers previously read as well as an interesting new one, which led them to iterate on their search query to find other papers by the same author.

However, some users expressed the desire for the keyword search features to support more robust or customizable queries. 
As a case in point, S01 conducted a global search for multiple keyword variations ``uncertainty visualisations'' $\rightarrow$ ``uncertainty visualizations'' $\rightarrow$ ``uncertainty visualization'', which returned 0, 8, and 49 hits respectively.
Fuzzy string matching would be a useful feature to support in subsequent iterations of \system{} (S06).
Another small usability issue that arose was lack of feedback upon clearing filters. 
For instance, some participants would backspace to delete text; however, the system would only remove filters by selecting the `x' icon next to the filter (S01, S05). 
Furthermore, S06 suggested it would be useful for \system{} to expand the searchable text beyond titles and abstracts: \textit{``Google Scholar searches body text too.''}

\subsubsection{Similarity Search}
\noindent\textbf{By Paper. } The ability to start with a seed paper(s) and identify other relevant literature was appreciated, with varying opinions about the quality and relevance of results.
Many participants were able to identify interesting and relevant papers; e.g., S04 identified a relevant paper from two key authors that they were not aware had collaborated.
S05 indicated a significant finding of a paper that \textit{``did something similar to what [they] were considering doing.''}
Compared to searching by keywords, S05 said \textit{``the papers [they are] seeing now are a lot more relevant. Some of these papers [they have] been reviewing. Some of them are kind of new.''}
S05 later acknowledged the utility of the similarity score: \textit{``It seems reasonable… Those on top tended to be more relevant to what [they were] looking for.''}
S06 commented that the similarity score was good feedback on the precision and quality of the search itself: \textit{``some would return like 0.0001 and [they] could see that [their] search was wrong.''}

Not all feedback about the similarity search was positive, however. 
S01 was uncertain about the quality of the results, stating they \textit{``could find a few papers that came up that slipped [their] mind, but [they] didn’t find any new papers that [they] hadn’t already cited. [...] [they] have some confidence that it would work, but for this particular context, [they] did not find anything new.''}
In response to some search queries, participants expressed disappointment with the results. 
For instance, using a single seed paper as input to similarity search, S02 indicated \textit{``these do not seem to be good results. The 2-D search does not seem to be good with GloVe. The N-D results were much better.''}
S02 then added additional papers as input to the similarity search and again noted \textit{``some match, but some do not. [...] [they] could have expected better search results.''} 
S02 ultimately suggested to explore other transformer embeddings, e.g., BERT.


\smallskip
\noindent\textbf{By Abstract. } While not all participants had an abstract prepared to utilize the Similarity Search by Abstract feature, they nonetheless saw value in it. 
S01, for instance, indicated that if they are \textit{``starting a new project [...] [they] can write up some words in the form of an abstract to see if this has been done.''}

S06 interestingly appropriated the abstract search in response to perceived shortcomings of traditional search features. 
For instance, after searching by keyword, applying filters, and iteratively revising queries to try to capture multiple keywords, S06 felt dissatisfied with the limitations of searching by keyword in \system: \textit{``Maybe [they] should use word embeddings because it might have more flexibility, and [they] can pass more information in [their] search.''}
They wrote a quick abstract paragraph during the study session and observed that the results showed \textit{``a lot of foundational literature.''}
They iterated, adding additional details to the abstract and expressed \textit{``Wow, this shows much better results now than the short abstract.''}
By the end of the study session, S06 identified several papers they had already cited as well as a few key new ones: \textit{``For 15 minutes, [they] found two papers [they] might be interested in. It’s a really useful process. Otherwise [they] might spend a lot of time scanning PDFs, which is not a very pleasant experience.''}

\subsubsection{Visualization Canvas}
Many participants found the projection visualization of the embedded space to be a useful way to identify conceptually ``nearby'' relevant papers. 
S05 suggested the visualization \textit{``provides a nice overview of the selected papers, and [they] could see to drill down into more details or look for clusters.''}
S01 appreciated the ability to select nearby papers in the embedded space via lasso, indicating \textit{``It's like a mystery. [They] feel like if [they] spent some time on this, [they] might stumble upon a paper that was relevant that was published in a different domain [...] It might be especially useful if [they] worked on a different topic that [they] had not worked on in the past.''}
S04 echoed this sentiment and added that the feature to locate a given paper on the visualization was helpful for orienting.

However, this impression was not universal. 
While S06 appreciated searching by abstract, they preferred to examine results in tabular format, because \textit{``personally [they are] not super familiar with these visualizations, dimensionality reduction, so it’s harder to interpret how to assess this information.''}
S03 was skeptical about the accuracy of the projection, stating \textit{``the algorithm might be bad, or the projection. It doesn't accurately depict similarity between papers.''}

Some participants suggested variations, such as spacing out papers in the visualization and connecting them by edges where the weight reflects the similarity with other papers in the visualization (S04).
S06 suggested for lasso selection, it would be useful to see \textit{``factors that can cluster similar papers.''}
Furthermore, S04 suggested additional interactivity to filter out papers on different ``layers'' in the visualization, e.g., those that are part of similarity search, saved papers, etc. 
S02 suggested a minor tweak: \textit{``when [they] do this similarity search, it should automatically zoom to show the paper(s) that were the beginning search point and the papers that it found, rather than this zoomed out view where [they] have to look for the orange or red dots.''}

\subsubsection{Meta View}
The Meta View went relatively unused compared to other features of \system.
However, some participants did express ideas to improve its utility. 
For instance, S03 expressed that they would have preferred if the Meta View \textit{``[did not display] keywords for the stuff above [Paper Collection View], but for what [they] have selected [Similarity Search input, Saved papers].''}
S05 suggested that the Meta View could offer additional keyword \emph{recommendations} based on semantically similar keywords, to help users identify other potential search terms. 
Others indicated a desire for further integration of the Meta View such that selecting a keyword could highlight papers in the visualization (S04) or filter the Paper Collection View (S06).

\subsubsection{Saved Papers Cart}
The Saved Papers Cart was also not used as often as some of the other views. 
Some preferred their existing workflow of downloading PDFs directly (S06), while others appreciated the ``cart'' analogy and the accompanying mindfulness to \textit{``fill the cart with relevant papers''} (S02) as an alternative to manually maintaining \textit{``a word document to keep track of the titles''} (S03). 

\begin{figure}[!t]
    \centering
    \setlength{\abovecaptionskip}{4pt}
    \setlength{\belowcaptionskip}{-10pt}
    \includegraphics[width=\columnwidth]{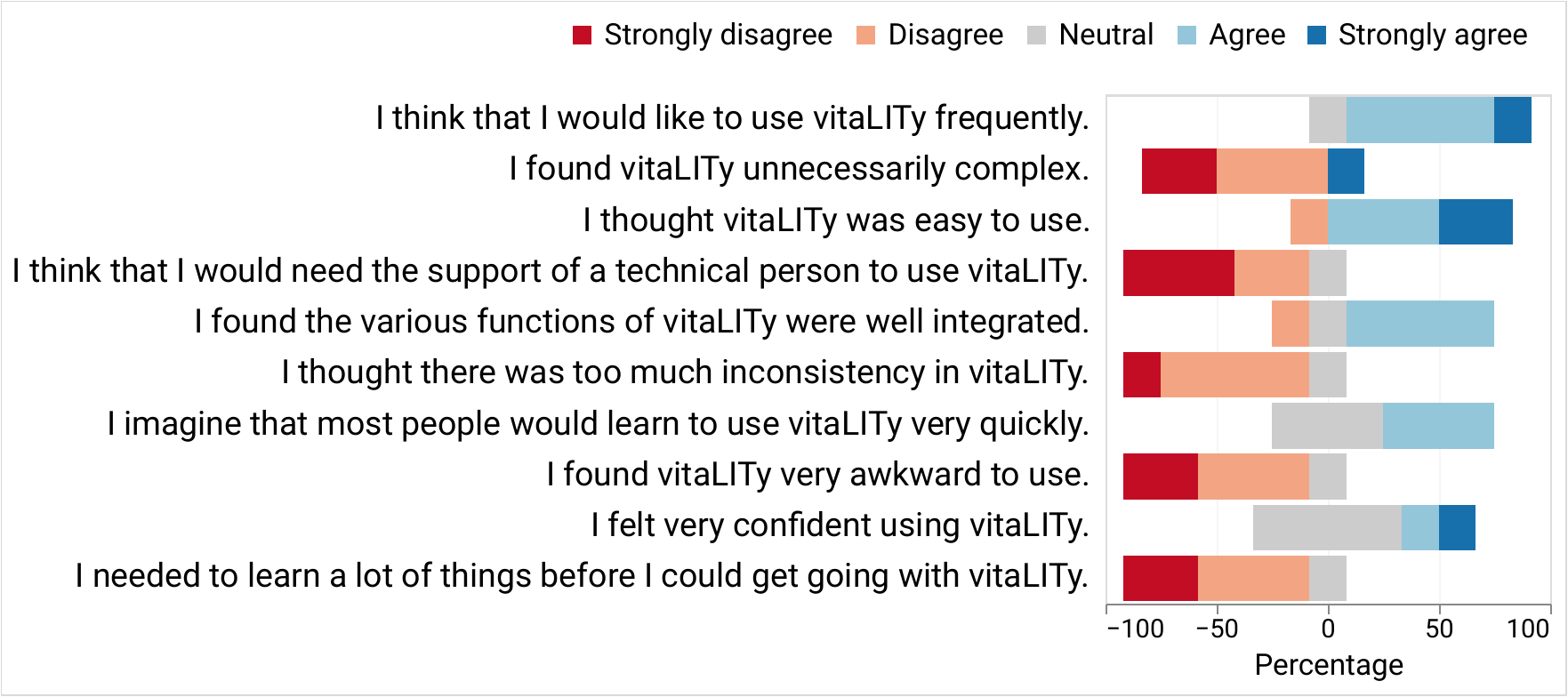}
    \caption{Overall SUS scores of \system{}.}
    \label{fig:sus-scores-generic}
  \end{figure}

\subsubsection{Summary \& Workflow}

\noindent\textbf{Overall Impressions. }
Users believed that \system{} would be useful in a variety of contexts.
Several users believed \system{} would be helpful in identifying gaps in their literature review (S01, S04). 
For instance, S04 indicated \textit{``it's very helpful to actually find a set of papers that are semantically relevant to one paper. If [they] identify a paper that [they] missed in the lit review, [they] can find other papers similar to that one to be sure [they] don't miss anything else.''}
Participants felt that it could help avoid ``embarassment'' of reviewers pointing out missing related work (S01, S06). 

\revised{The individual SUS scores per participant were S01=72.5, S02=77.5, S03=45, S04=72.5, S05=70, S06=92.5 for an overall average SUS score=72.5 (Figure~\ref{fig:sus-scores-generic}).} While participants generally liked using \system{}, several expressed that, given the large number of features, customization of the screen real estate would have been beneficial (S02, S04, S06).
For instance, S02 indicated \textit{``when [they] had already filtered by keywords, [they are] only focusing on this view [Visualization Canvas]. It’s very small in the screen space. [They] want to hide the Meta View and maybe even the [Paper Collection View], so [they] can easily zoom and pan and lasso. The Similarity Search panel could also be bigger.''}
Others echoed formative feedback, wanting to see the citation network in \system, e.g., which papers cite others (S01, S06).

\smallskip
\noindent\textbf{Workflow. }
Some participants viewed \system{} as a complementary component to their existing literature review workflow. 
For instance, S01 indicated they would \textit{``interleave this with a Google Scholar search. If [they] found a few relevant papers, [they] would go to Google Scholar to see the references in that paper and who has cited that paper.''}
S06 indicated preference to continue their existing approach of beginning a literature review with Google Scholar and use \system{} at a later stage of the research, e.g., \textit{``when [they] want to do some sanity checks [...] [after they] have [their] abstract, papers [they] have already cited, and based on that [they] can do a more narrow search for papers [they] might be missing,''} while others preferred to use \system{} as early in the lit review process that you are able to \textit{``structure the related work sections [...] and [identify] those 2-3 themes''} (S05).

Others felt that \system{} suffered from many of the same problems that existing tooling has. 
For instance, S01 said \textit{``[The] target is one unknown paper among hundreds. A lot of the papers [they] find because coauthors tell [them] about them.''}
S03 indicated they would use the tool primarily in the same ways as Google Scholar, e.g., \textit{``[they] would just search for keywords.''}


%% file: sections/discussion.tex
\section{Discussion}
\label{sec:discussion}

\medskip
\noindent\textbf{Quality of Search Results. }
Across our (relatively small) sample of participants, there was variability in terms of perceived relevance of Similarity Search results. 
Some participants felt that, like Google Scholar, relevant results were lost among a sea of irrelevant papers, while others felt that the results were highly relevant. 
In general, participants perceived results from SPECTER embeddings to be more relevant than GloVe, suggesting that further exploration of alternative transformer-based approaches (e.g., BERT~\cite{devlin2018bert}, or training a custom model on the target document corpus) could yield better search results. 
Furthermore, given the disparity in perceived quality and disparity in participants' perception of when this approach could be useful in their literature review process, future work could develop additional guidelines that assess the specific role of document retrieval based on semantic similarity.

\medskip
\noindent\textbf{Relevance \& Space. }
Presuming \system{} is able to provide serendipitous discovery of relevant literature, the process doesn't abruptly come to a successful end. 
Authors still need to manage goals in their writing that may be at odds with one another: i.e., the tradeoff of relevance or salience of related work and the commodity of space. 
From this perspective, \system{} is best viewed as a way to identify critical gaps or serve as kindling for a new literature review. 
In its current form, \system{} shows (1) similarity score, and (2) citation counts as the primary cues of relevance or salience of a given paper.
It still requires substantial knowledge from the author to (1) read an abstract or paper and determine its actual relevance to a given topic, and (2) assess the credibility of the work, author(s), and venue. 
Subsequent versions of \system{} could focus on innovating solutions to support these and other parts of the literature review process.

\medskip
\noindent\textbf{Future Work. } 
Based on our use of \system{} and participant feedback, we identify a number of potential future directions. \revised{First, as mentioned in the Related Work, with citation and user activity data, \system{} could expand its functionality to citation or read/view recommendation using SPECTER.}
\revised{Second}, current similarity scores in the projected 2-D space (UMAP) are based on the reciprocal of the distance measure and might yield different results compared to distances in the N-D embedding space. 
These scores and their context may not be especially intuitive for users. 
Hence, future work could refine the similarity score formulation and / or presentation in \system{} to provide users an accessible framework to interpret results.
\revised{Third}, the Saved Papers Cart currently exports a file in JSON format with the papers. 
At least two improvements could be made within this view, including exporting files in .bibtex format for easy incorporation in \LaTeX{} bibliographies.
Furthermore, it could be useful to users to provide a meta analysis of the saved papers, e.g., via topic modeling. 
How can these papers be summarized? 
\revised{Fourth, while our research prototype of \system{} is intended to be complementary to existing search strategies, future work could expand \system{} to a more comprehensive search tool, incorporating the benefits of e.g., citation networks.}
\revised{Lastly, \system{} is modular, scalable, and extensible: it applies the virtual scrolling principle in the UI table views (preventing unnecessary rendering of objects not visible in the viewport), renders the UMAP using WebGL, and uses a library (faiss) that performs efficient similarity search of dense vectors with an option to leverage GPUs. The \emph{scraper} module currently uses DBLP as the source of raw data but can be extended to support other digital libraries, e.g., JSTOR (https://www.jstor.org/).
Hence, augmenting the system with additional venues (and allowing users to define which venues are relevant to load in their specific literature review) is a feasible next step to expand \system{} to other research domains.}

%% file: sections/conclusion.tex
\section{Conclusion}
\label{sec:conclusion} 

We introduced a visualization system, \system, designed to promote serendipitous discovery of relevant academic literature. 
Designed and developed with formative input from data visualization researchers, \system{} allows users to search and explore academic literature using a document-level transformer-based approach to identify semantically similar literature.
In addition, we contributed a dataset about \texttt{59,232} academic articles with metadata (titles, abstracts, authors, keywords, citation counts, etc.) across \texttt{38} venues common in data visualization research, along with open-source scrapers to expand and customize the corpus of literature searchable in \system.
We demonstrated how \system{} can complement existing academic literature review practices through a series of usage scenarios and shared feedback from 6 data visualization researchers from a qualitative study. 
Participants expressed excitement to incorporate \system{} in their workflow, to identify gaps in their academic literature searches or to kickstart the literature review of a new topic. 
\revised{While our initial prototype and evaluation focused on the data visualization field, we have open-sourced our system and scraper framework to enable expansion of the \system{} approach to other venues and academic communities.
We invite those who are interested to augment the \system{} system and data for their academic interests.}